# Multicaloric effect in Fe$_{48}$Rh$_{52}$ alloy: case of combination magnetic field and uniaxial tension

Abdulkarim A. Amirov[1]*, Dibir M. Yusupov[2], Alexey S. Komlev[1,3], Maksim A. Koliushenkov[1,3], Alexander P. Kamantsev[4], Akhmed M. Aliev[2]

[1]*National University of Science and Technology MISiS, 119049, Moscow, Russia*

[2]*Amirkhanov Institute of Physics of Dagestan Federal Research Center, Russian Academy of Sciences, 367003, Makhachkala, Russia*

[3]*Physics Department, Lomonosov Moscow State University, Leninskie Gori 1, 119991 Moscow, Russia*

[4]*Kotelnikov Institute of Radioengineering and Electronics of Russian Academy of Sciences, 125009 Moscow, Russia*

* Corresponding author, E-mail: *amiroff_a@mail.ru*

**Abstract.**

Mono and multicaloric effects in Fe$_{48}$Rh$_{52}$ alloy under applied magnetic field, uniaxial tension and their combination were studied by direct method. It was found that for single cases, the inverse caloric effect was observed with $\Delta T_{AD}$ = -2.9 K (1T) at 330 K in the case of the magnetocaloric effect and $\Delta T_{AD}$ = -0.5 K (104 MPa) at 328 K in the case of the elastocaloric effect. The combination co-application of the external 1 T magnetic field and a 104 MPa tensile results to the observation of a synergistic effect with $\Delta T_{AD}$ = -3.4 K at 330 K when a, which exceeds similar values for mono caloric effects. As was shown from comparison of calculation and experiments multicaloric effect it is not a sum of mono caloric effects and several factors as geometry of the sample as well the protocol for applying external fields should be taken into account. It was shown that the distribution of mechanical stresses in the Fe$_{48}$Rh$_{52}$ sample with a geometry in the shape of a plate with holes is heterogeneous, which should be taken into account when measuring calorific effects using tension through holes

Multicaloric materials exhibit a multiferroic phase transition that can be induced by diverse external fields, and that the caloric effects that arise from the sequential or simultaneous change of more than one external field are the multicaloric effect (MultiCE), which offer promising perspectives for the development of solid-state cooling technologies (CE) [1–4].

The family of near-equiatomic FeRh alloys ordered to *B2* type crystal structure with a volume-centered cubic lattice (type *CsCl*) is considered as one of the benchmark materials exhibiting giant CE and multiCE [5]. The *B2* ordered FeRh alloys have a coexistence of ferromagnetic and ferroelastic ordering and exhibits a metamagnetic transition from a low-temperature antiferromagnetic (AFM) phase to a high-temperature ferromagnetic (FM) phase, in which an isotropic expansion of the crystal lattice occurs by $\Delta V/V$ ~1% without changing its symmetry [6]. The uniqueness of the functional properties of the FeRh alloys lies in the fact that they exhibit «giant» magnetocaloric (MCE) [7], barocaloric (BCE), elastocaloric (ElCE) [8] and multicaloric (MultiCE) effects [9–11] at room temperature region. Moreover, the simple symmetry of the crystal lattice, room temperature of phase transition (PT) and changes their magnetic properties around PT temperatures makes them good model objects for studies the phenomena related with a first-order magnetic phase transition (FOMPT).

MultiCE at combined magnetic field and hydrostatic pressure was experimentally studied in $Fe_{49}Rh_{51}$ alloy using magnetic measurements at applied magnetic field and hydrostatic pressure [9–11]. The MultiCE under uniaxial compression and a magnetic field using calorimetric measurements were studied by Gràcia-Condal et al [12]. In these works, MultCE were studied by indirect methods and multicaloric studies using direct measurements have challenges.

The motivation of our work is conducted to the studies of MultiCE in FeRh alloys by direct methods at two co-applied external fields, when the increase of total CE is expected. It requires solving the following main problems: (i) development of experimental methods and construction of experimental setup for direct measurements of MultiCE; (ii) searching of optimized protocol of co-application of external fields, (iii) investigation of some side effects (pre-history of the sample and its shape, role of interrelations between magnetic and structural sub-systems). As one of the prospective multi-stimuli approach the applying of an external magnetic field and uniaxial tension was considered.

For studies of MultiCE, the sample from an alloy ingot of the composition $Fe_{48}Rh_{52}$ fabricated by induction melting from high-purity powders of Fe (99.98%) and Rh (99.8%) elements in an argon atmosphere was used, the magnetic and structural properties were reported earlier [13]. According direct MCE measurements, the $Fe_{48}Rh_{52}$ sample exhibits AFM-FM transition temperature $T_m$ in the region of ~314 K in a magnetic field of 0.62 T in heating protocol [14].

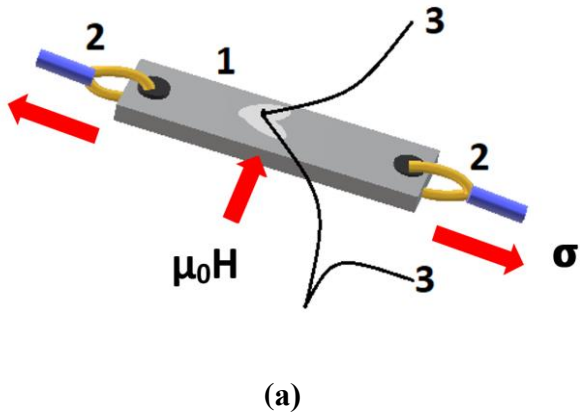 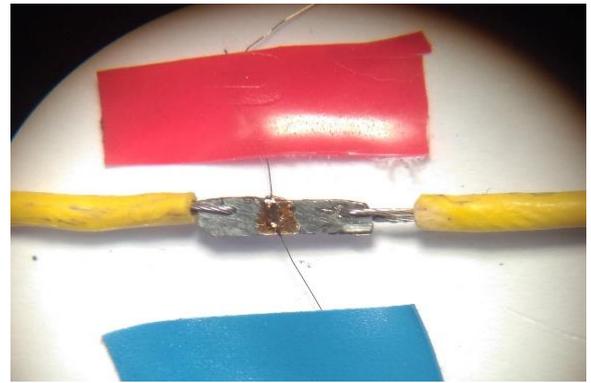

**(a)** **(b)**

*Figure 1. A scheme of direct MultiCE measurements (1 – sample, 2 – ropes, 3 – differential thermocouple) (a) and a photo of one of the mounting options for the $Fe_{48}Rh_{52}$ sample, using a steel rope and a chromel-constantan thermocouple (b).*

For our experiments, the $Fe_{48}Rh_{52}$ sample was flattened to a thin plate with a thickness 0.32 mm and then was cut in the shape of a rectangular with dimensions 5*1.7 mm (Length×Width). Holes with a diameter of about 0.5 mm were drilled at the edges of the sample (1) for pulling rops (2), with which, according to the scheme shown in Figure 1a, the sample was stretched uniaxially. A microthermocouple (3) (chromel-constantan or copper-constantan) prepared by spot welding was glued to the center of the sample. The MCE, ElCE and MultiCE were studied using direct method, by measuring of the adiabatic temperature change $\Delta T_{AD}$, when mono (for ElCE or MCE) or simultaneously (for MultiCE) external stimuli (magnetic field and uniaxial tension) was applied. The ElCE was measured based on the method described in [8], where uniaxial tension was applied using a metallic rope attached to the sample through holes drilled in it and thrown through a mechanical block, to the end of which loads of various weights were suspended. For MultiCE measurements, this experimental setup was upgraded by installation of 1 T magnetic field source with a Halbach structure mounted on linear displacement system (*Figure 1 in Supplementary Information*). The magnetic field was applied in plane of the sample and perpendicular to the axis along which uniaxial tension was applied (Figure 1a.). For test experiments, the different mounting options for thermocouples and ropes were used, without changing of general mounting scheme: the case with using of a steel rope and a chromel-constantan microthermocouple is shown in Figure 1b. Preliminary ElCE experiments have shown that metallic rope leads to significant errors due to heat loss on the rope, the mass of which significantly exceeds the mass of the sample and has good thermal conductivity. To solve this problem, a 0.5 mm thick Kevlar rope was used for further experiments. Final measurements were carried out using a differential copper-constantan thermocouple made of copper and constantan wires with a diameter 40 μm. To reduce the parasitic EMF, the microthermocouple wires were twisted together. EMF signal from microthermocouple was measured by multimeter (Keithley 2000). To control the desired temperature, the LakeShore Model 335 temperature controller was used.

Figure 2 shows the temperature dependences of the magnetization *M(T)* of the initial (Figure 2a) and flattened (Figure 2b) sample $Fe_{48}Rh_{52}$ at different applied magnetic fields.

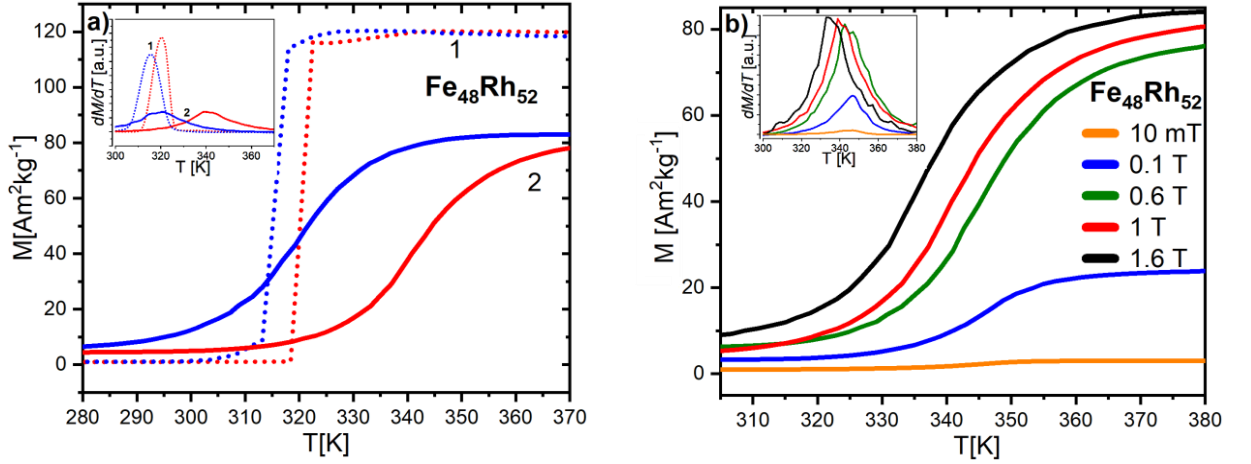

*Figure 2. Temperature dependences of magnetization of the initial (1) and flattened plate (2) of the $Fe_{48}Rh_{52}$ sample prepared for ElCE measurements (a); temperature dependences of the magnetization of the $Fe_{48}Rh_{52}$ sample in the form of a plate (2) at various values of applied magnetic fields up to 1.6 T (b); magnetization derivatives at temperature dM/dT for M(T) measurements (inserts).*

As can be seen from *M(T)* curves for the initial and mechanically flattened $Fe_{48}Rh_{52}$ samples, mechanical action leads to the degradation of its initial magnetic properties: magnetization in the FM state decreases from ~120 $A*m^2*kg^{-1}$ to ~83 $A*m^2*kg^{-1}$, and the temperature of the AFM-FM transition, estimated from the temperature derivative of magnetization *dM/dT*, shifts towards high temperatures by ~ 20 K. The increase in the phase transition temperature during deformation of the sample is explained by the presence of residual quenching stresses in the sample. In this case, the AFM-FM and FM-AFM transition for the flattened $Fe_{48}Rh_{52}$ sample is observed in a wider range with a temperature hysteresis width of about ~ 20 K, while for the initial sample a sharp change in magnetization was observed in the transition region with a limited temperature range with a temperature hysteresis width of about 4.5 K. There are various hypotheses regarding the causes affecting changes in the width of the temperature hysteresis, which are discussed in [15,16]. As seen the mechanical stress leads to increase non-zero magnetization of the sample in the low-temperature state (Fig.2a). As possible mechanism we suggest it with increase in the proportion of residual FM phase on local macroscopic crystalline defects and to explain in the additional measurements of magnetization were carried out (*Figure 2 in Supplementary Information*).

The analysis of the *M(T)* dependences measured at different magnetic fields for the flattened $Fe_{48}Rh_{52}$ sample allows to determine the shift of the transition temperature $dT_m/\mu_0 dH$=-6 K/T. The obtained value is comparable to $dT_m/\mu_0 dH$ =-7.2 K/T, observed in $Fe_{49}Rh_{51}$ alloy [17]. The parameters of PT in the sample prepared for multicaloric studies were also estimated from calorimetric measurements (specific heat $C_p$ vs temperature) as maxima of $C_p(T)$ in a zero and 1 T magnetic field and corresponds to temperatures of AFM-FM (FM-AFM) 330.6 K (338.6 K) in a zero and 321.2 K (329.2 K) at 1 T magnetic field with temperature hysteresis ~8 K (*Figure 2 Supplementary Information*). The shifts of the transition temperature were $dT_m/\mu_0 dH$=-9.5 K/T (AFM-FM), which is closer to the experimental values of -9.7 K/T for $Fe_{49}Rh_{51}$ and -9 K/T for $Fe_{48}Rh_{52}$ calculated from calorimetric measurements [9,18].

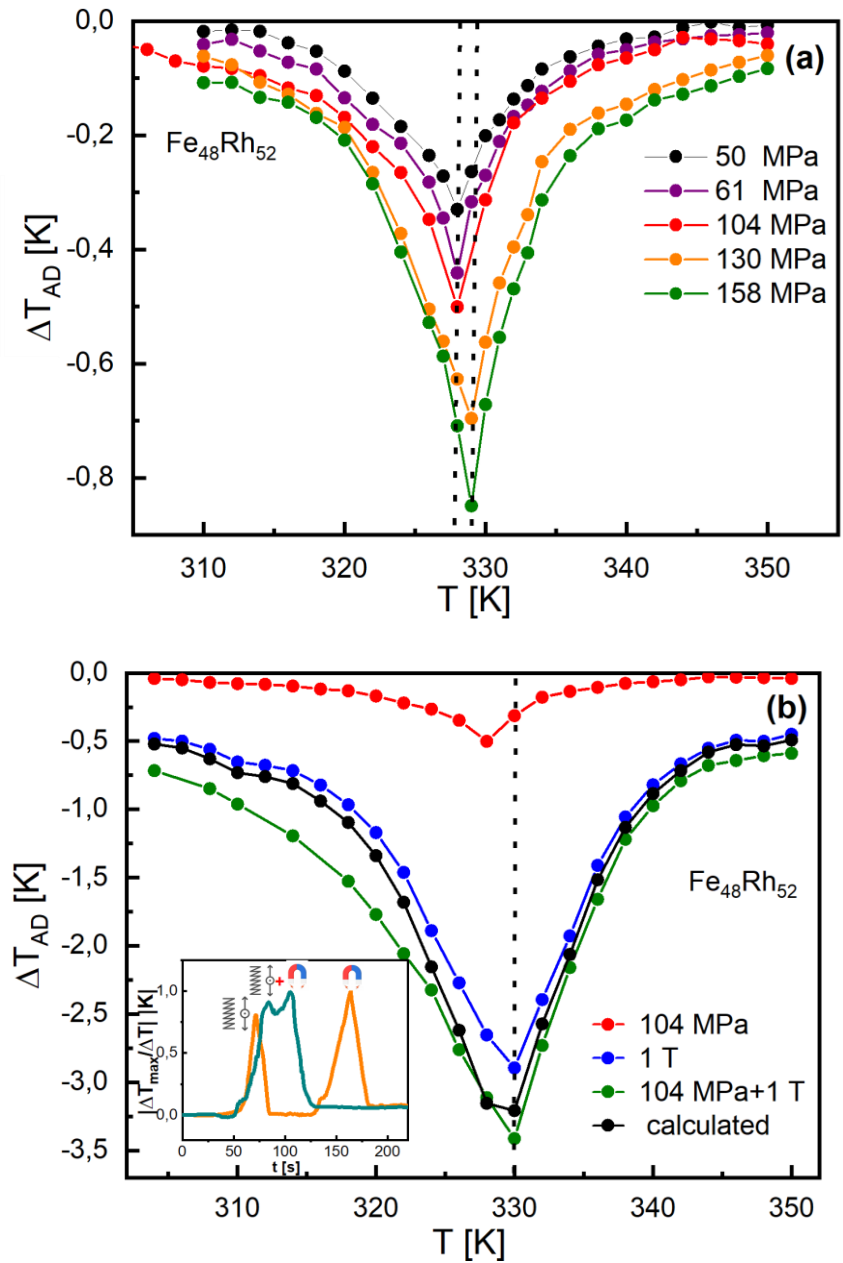

*Figure 3* Temperature dependences of the adiabatic temperature change $\Delta T_{AD}$ for ElCE (a) at selected values of uniaxial tension and MultiCE with measurements of MCE at 1 T magnetic field and ElCE at 104 MPa tension (b). Time profile of normalized $\Delta T_{AD}$ with differ time shifts

*between applied magnetic field and tension. The maximums corresponded of each type of external fields are marked by icons of magnetic field and uniaxial tension.*

Figure 4a shows the results of direct ElCE measurements for the $Fe_{48}Rh_{52}$ sample carried out at different mechanical stress applied in uniaxial tension mode. The temperature dependences of the adiabatic temperature change $\Delta T_{AD}$, measured in the heating at various values of tensile mechanical stress, show maxima in the temperature range corresponding to the AFM-FM transition, which slightly shift towards high temperatures with an increase in the magnitude of the mechanical stress. One should note that, according to the literature data, application of uniaxial strain shifts the PT temperature of FeRh towards lower temperatures with a rate of $dT_m/d\sigma$ ~-20 K/GPa [8], while application of hydrostatic pressure shifts it towards higher temperatures with a rate of $dT_m/dP = 54$ K/GPa [9]. In our case, the rate of PT temperature shifting estimated from the maximum of $|\Delta T_{AD}|$ has a positive value and is $dT_m/d\sigma=10$ K/GPa in the high temperature region for the AFM-FM transition, which differs from the data $dT_m/d\sigma$ ~-20 K/GPa given in the literature [8]. For comparison, uniaxial compression stabilizes the AFM phase and shifts the temperature of the AFM-FM transition towards high temperatures and $dT_m/d\sigma =30$ K/GPa with a positive sign. This unusual behavior allows us to make the suggestion about inhomogeneous distribution of the deformation in sample, where the part of sample is stretched and part is compressed (for example, as in the case of twisting). It can be also suggested the situation, when the part of sample is transformed in one direction (FM-AFM) and another part in other direction (FM-AFM). In our experiments, a sample of small sizes relative to the cable was used, and the holes for attaching the cables are not fully symmetrically i.e. the centers of the holes were not on the same line running along the sample along its center (or shifted by different distances). Thus, the stretching can be not strictly uniaxial, and the resulting deformations are heterogeneous due to the possible twisting of the rope when it is stretched. The contradictions regarding the positive $dT_m/d\sigma >0$ and the observation of the inverse ElCE under uniaxial tension are also an indirect sign of the occurrence of inhomogeneous tensile deformations in the sample. Estimations of the ElCE parameters using the Clapeyron-Clausius ratios in this case will be incorrect due inhomogeneous tensile deformations in the sample conditions. For example, the 59% difference between the experimentally measured and calculated values of $\Delta T_{AD}$ using the Clapeyron-Clausius ratios was observed, which was associated with non-isothermal transition conditions and possible inhomogeneous deformations in a real sample [8]. Anyway, more extensive studies are needed to explain the nature of real stresses in sample.

Next, experiments were conducted to direct measurements of the multiCE in which the adiabatic temperature change $\Delta T_{AD}$ was measured when the sample was stretched under 104 MPa and a magnetic field of 1 T was applied. The time of application of the magnetic field and tension was chosen experimentally and during the test experiments it was found that the better $\Delta T_{AD}$ is observed, when external fields were applied not simultaneously, but with a small-time shift of $\sim\pi/4$. This behavior is explained by the different dynamics of MCE and ElCE, at which the rates of change of the magnetic field and mechanical loads, as well as the rates of change $\Delta T_{AD}$ differ, similar differences were observed for another FOMPT materials: LaFeSi-based systems [19] and Heusler alloys [20].

Figure 3b shows the results of measurements of the temperature dependences of the $\Delta T_{AD}$ for mono (MCE and ElCE) and combined (MultiCE) caloric effects. For single cases, the inverse caloric effect is observed $\Delta T_{AD}^{MCE}$ = -2.9 K (1T) at 330 K for MCE and $\Delta T_{AD}^{ElCE}$ = -0.5 K (104 MPa) at 328 K for ElCE. As seen, combination of the external fields leads to observation of synergistic effect with $\Delta T_{AD}^{MultiCE}$ = -3.4K at 330 K, when 1T magnetic field and 104 MPa tension was co-applied. As seen from comparison of calculated $\Delta T_{AD}^{MultiCE}$ *(black points)* =$\Delta T_{AD}^{MCE}$*(blue points)* + $\Delta T_{AD}^{ElCE}$ *(red points)* and measured MultiCE (green points) it is not a sum of mono CE (MCE and ElCE). As were shown for a Ni-Mn-Ga-Co alloy the cross-coupled contribution is minimal, when the second applied field is small or close to zero, large effect is observed, when a both fields ( magnetic and mechanical) are large [21]. It should be note that cross-coupled contribution in our experiment depends on temperature: it is larger in 310-320 K - temperature region, where FeRh sample has initial PT temperature before flattering to plate shape sample. As was shown , general entropy change at co-applied uniaxial tension and a magnetic field is not sum of mono CE (MCE and ElCE) and contains the third term is responsible for the contribution from cross-effects (*Relation 1 in Supplementary information*) [12].

To explain the nature of real deformations occurred in sample under tension, calculations based on finite element method (FEM) using the COMSOL Multiphysics and experiments using strain gauges were carried out. For calculations the model with geometry corresponded to the real sample were built and the main parameters of the model used for calculations are collected in Table 1.

*Table 1. Parameters of the FeRh sample for calculations FEM calculations.*

| Sample dimensions (L×W×T ), mm | Diameter of hole, mm | The distance between the edges of the holes, mm | Density, g/cm$^3$ | Young's modulus, GPa | Poisson's ratio |
|---|---|---|---|---|---|
| 5*1.7*0.32 | 0.5 | 2.5 | 10 | 170[22,23] | 0.3 |

Calculations were carried out for two cases: symmetrical (Figure 4a-b) and asymmetric (Figure 4 c-d) hole arrangement. For the asymmetric case, the centers of the holes were shifted relative to each other by 100 μm. As can be seen, for both cases, the nature of deformations is heterogeneous, which is more pronounced in the case of asymmetric holes (Figure 4 c-d). The calculated model is close to the well-known problem of mechanics corresponded to the stretching of finite and infinite plates with a circular hole (Kirsch problem). In problems of uniaxial stretching of plates with a hole (or several holes), the localization of inhomogeneities is observed around the holes and depends on various factors (the shape of the hole, the ratio of the size of the hole and the thickness of the plate, the length of the plate, etc.) [24]. Thus, the presence of holes is one of the factors influencing the heterogeneous distributions of mechanical stresses, and in the case of displaced holes they are more pronounced.

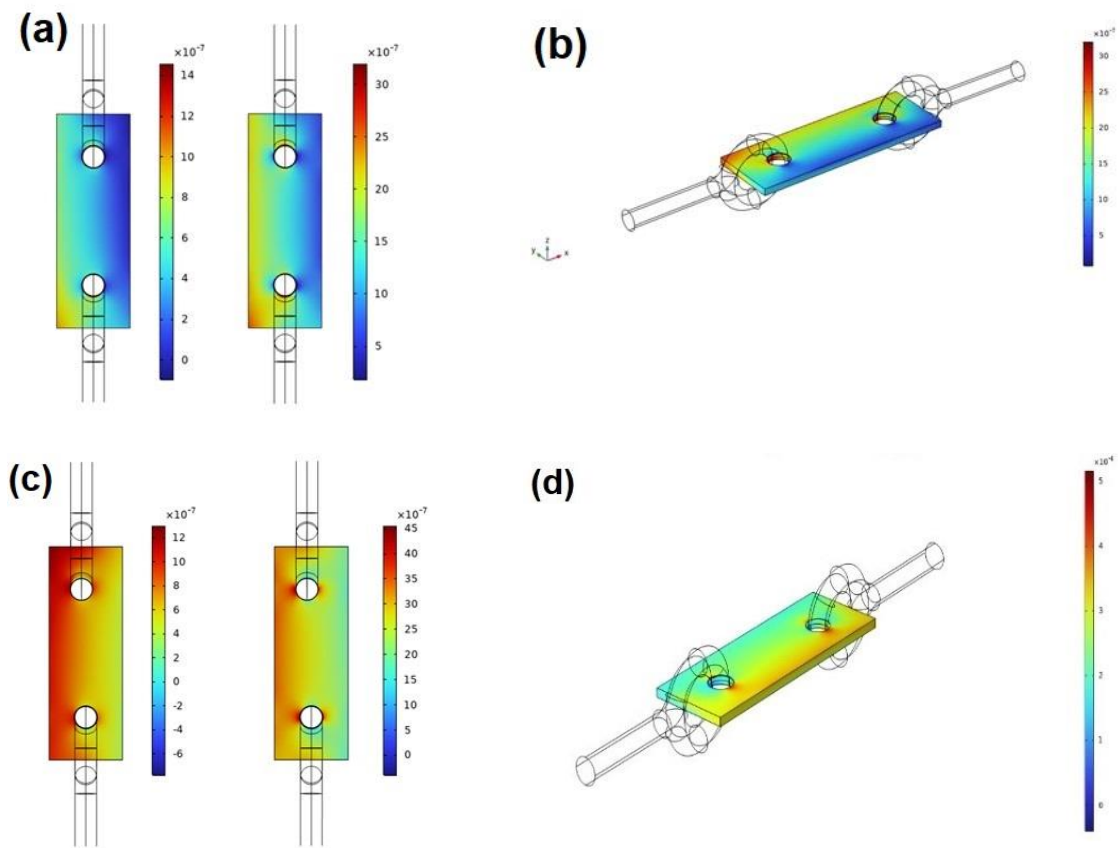

*Figure 4. Calculations of deformation at uniaxial tension of the $Fe_{48}Rh_{52}$ sample in the case of symmetrical (a-b) and asymmetric (c-d) arrangement of holes for the cable at a tensile mechanical stress of 158 MPa; (a) and (c) -2D distribution of deformation from upper and lower side of the sample, (b) and (d) - 3D distribution of deformation.*

To verify the calculations and confirm the presence of heterogeneity in the distribution of mechanical stresses in the sample during its tension, which occur under the conditions of our experiments, experiments were conducted to measure the relative linear change ε of the sample at various tensile stresses σ at fixed temperatures. For this purpose, two strain gauges were attached to the upper and lower sides of the sample, according to the scheme shown in the insert to Figure 5.

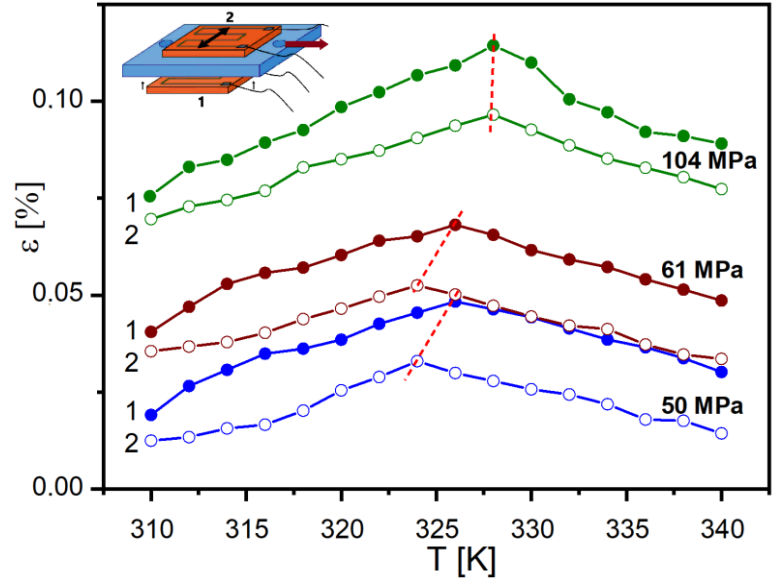

*Figure 5. Temperature dependences of relative linear changes ε at different values of tensile stresses (scheme of mounting of strain gauges 1 and 2 shown in the insert).*

Strain gauges *1* and *2* measured the deformation in only one direction (orthogonally to direction where stress is applied, according scheme of mounting of strain gages as shown in the insert to Figure 5) and the main purpose of the experiment was to observe the presence of asymmetry of deformations between the upper and lower sides of the sample plate. The temperature dependences of the elongation $\varepsilon=\Delta l/l_0$ ($l_0$ -is the initial size of the sample) at different values of tensile stresses measured from both strain gauges are shown in Figure 5. The values of ε are expected to demonstrate a maximum in the AFM-FM transition region, where the large volume change for *B2* ordered FeRh structure is observed. As it can be seen, the obtained values of strain gauges 1 and 2 are different, although they have the same sign. This allows us to conclude that under inhomogeneous deformations throughout the sample a uniaxial stretching is experimentally observed. An inhomogeneous character of deformations was observed in [25] for two-layer FeRh/PZT composite in result of applied magnetic field and electric field induced strain sample was bent. Strain gauge measurements were carried out according to a similar protocol and the values of ε from both gauges had different signs, which corresponded to the bending of the composite. The Young's modulus *E* estimated from strain measurements depends on the temperature, at the transition region it has a value of ~ 100 GPa and due to the heterogeneity of deformations are rather rough. In [26], the Young's modulus and internal friction for the $Fe_{49}Rh_{51}$ alloy in the AFM-FM transition region were investigated, they depend on the temperature, measurement cycles, and heat treatment conditions of the sample, and in the AFM-FM transition region *E~250 GPa* was observed in one of the measurement cycles. The maximum value of $\Delta T_{AD}$ =-0.85 K at 158 MPa is observed at a temperature of 329 K, which is significantly less than the value of -2.47 K at 150 MPa tension (-5.17 K at 529 MPa), which was observed for $Fe_{49}Rh_{51}$ [8]. The small values of the observed ElCE on our $Fe_{48}Rh_{52}$ may be attributed to the effect of nonelastic deformations, as a result plate shape, and to the non-uniform nature of elastic deformations due to the geometry of the sample and holes.

As a conclusion, the MultiCE in a FeRh sample with the shape of a plate with holes at co-applied magnetic field and uniaxial tension, it is not a sum of mono CE and to enhance the total caloric performance several factors should be take into account: 1) geometry of the sample and 2) the protocol for applying external fields. It will allow to achieve the synergetic effect in multicaloric measurements and will open new prospective for development of multicaloric cooling principles.

**Acknowledgments**

This research was supported by the Russian Science Foundation (project no. 24-19-00782, https://rscf.ru/en/project/24-19-00782/). Authors are grateful to Dr. Adrià Gràcia-Condal (Grup de Caracterització de Materials, Departament de Física, EEBE and Barcelona Research Center in Multiscale Science and Engineering, Universitat Politècnica de Catalunya, Eduard Maristany, 10-14, Barcelona, 08019, Catalonia, Spain) for fruitful discussions and recommendations.

**References.**

**Supplementary Information**

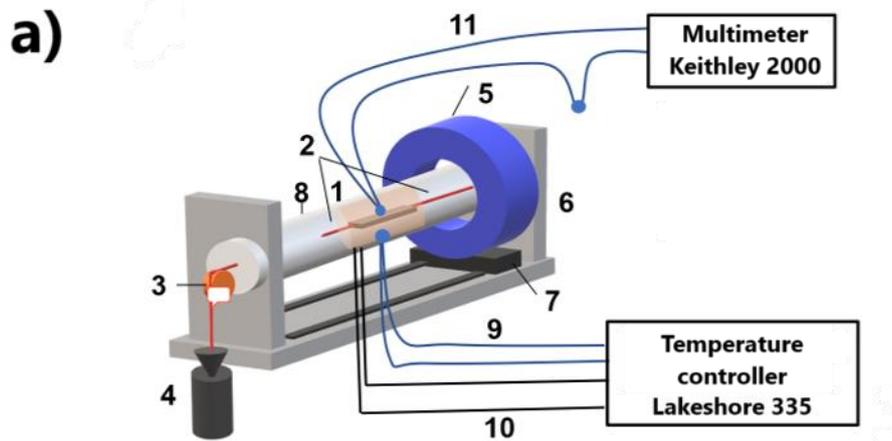

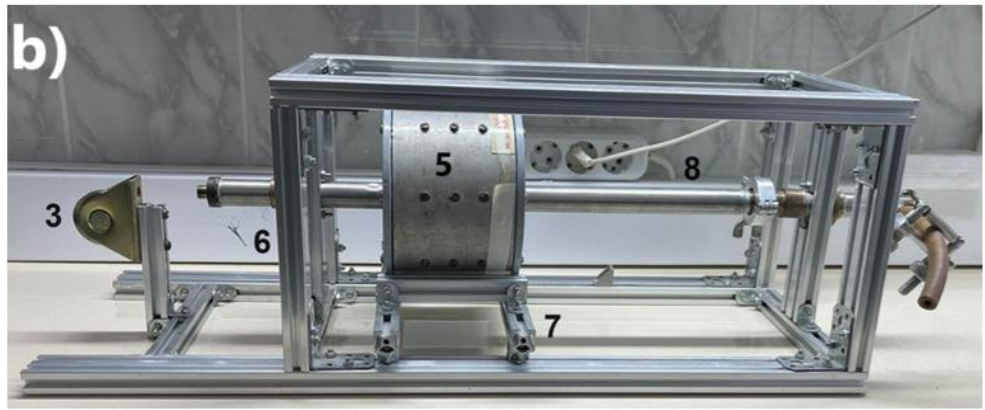

*Figure 1.* Scheme (a) and photo (b) of an experimentalsetup for direct measurements of MCE, ElCE and MultiCE at various regimes of applied magnetic field and uniaxial tension: 1-sample, 2-rope, 3-block, 4-load mass,5-Halbach type magnetic field source,6-bearing frame,7-roller system for moving the magnet, 8- adiabatic chamber,9-thermocouple,10-heater.

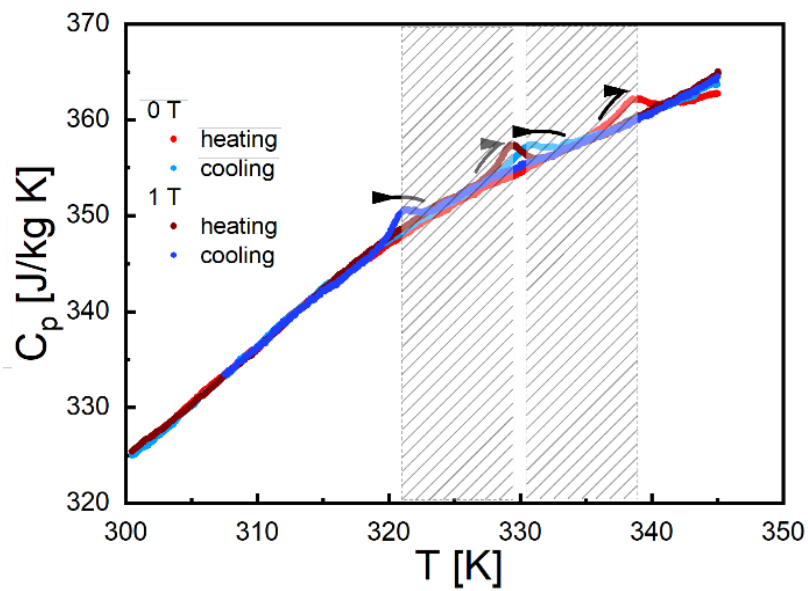

*Figure 2.* Temperature dependences of the specific heat in the heating and cooling modes for the $Fe_{48}Rh_{52}$ sample at a zero and 1 T magnetic field.

It is worth paying attention to the non-zero magnetization of the sample in the low-temperature state for both $Fe_{48}Rh_{52}$ samples, which increases in the sample after mechanical action. This is probably due to an increase in the proportion of residual FM phase on local macroscopic crystalline defects. Degradation of the parameters of the magnetic phase transition for $Fe_{48}Rh_{52}$ (decrease in magnetization in the FM state, increase in the width of the temperature hysteresis, change in the differential characteristics of the temperature dependence of magnetization) is primarily associated with irreversible effects resulting from inelastic mechanical deformation of the sample. In the areas of occurrence of mechanical stresses, there is a lattice deformation (local change in lattice parameters), where FM clusters are formed, which exist at low temperatures due to the dominance of FM exchange interaction in the discussed areas. In [27,28] a phenomenological model was proposed describing the origin and evolution of the growth of the FM phase in the field of MFPs of the first kind based on the Bean-Rodbell and Kolmogorov-Johnson-Mel-Avrami models using the example of a thin film $Fe_{49}Rh_{51}$. The origin of the FM phase begins on the surface of the $Fe_{49}Rh_{51}$ film, and then it grows over the entire surface. In this case, the PT ends with the growth of the FM phase near the interface between the alloy and the substrate, which indicates the dominant role of mechanical stresses in the processes of nucleation and growth of the FM phase. The presence of similar mechanisms of phase formation is possible in our case. From the results described above, it follows that in the studied sample there are FM clusters existing at low temperatures, in which tensile stresses prevail. In areas where compressive stresses are present, the transition temperature to the FM phase increases [29], which is confirmed by an increase in the temperature of the phase transition (Figure 2). On the other hand, near areas with local crystalline defects, mechanical stress gradients occur, and they can be quite stable with temperature changes. Therefore, the growth of the phase can still begin from the surface and in this sense, the activation energy of the nucleation of the phase is directly related to the surface energy.

In order to determine the nature of the growth of the FM phase inside the sample, hysteresis loops were measured at different temperatures during the heating process *(Figure 3 a)*. The temperature dependence of the coercive force and the squareness coefficient of the hysteresis loop $M_r/M_s$ are shown in Figure 2b. The high value of the coercive force in the low temperature region is explained by the presence of unidirectional exchange anisotropy due to the coexistence of AFM and FM phases. An additional confirmation of this fact is the displacement of the hysteresis loop relative to the zero field. The general tendency to decrease the coercive force is explained by a decrease in the volume fraction of the AFM phase during heating. It is noteworthy that at a temperature of 340K, the coercive force becomes less than 50 Oe. But according to the temperature dependences of magnetization, there is no significant increase in magnetization at this temperature.

This means that the nucleation process most likely prevails at the beginning of the evolution of the phase transition. Further heating leads to the dominance of the processes of growth and unification of FM clusters, as it usually happens.

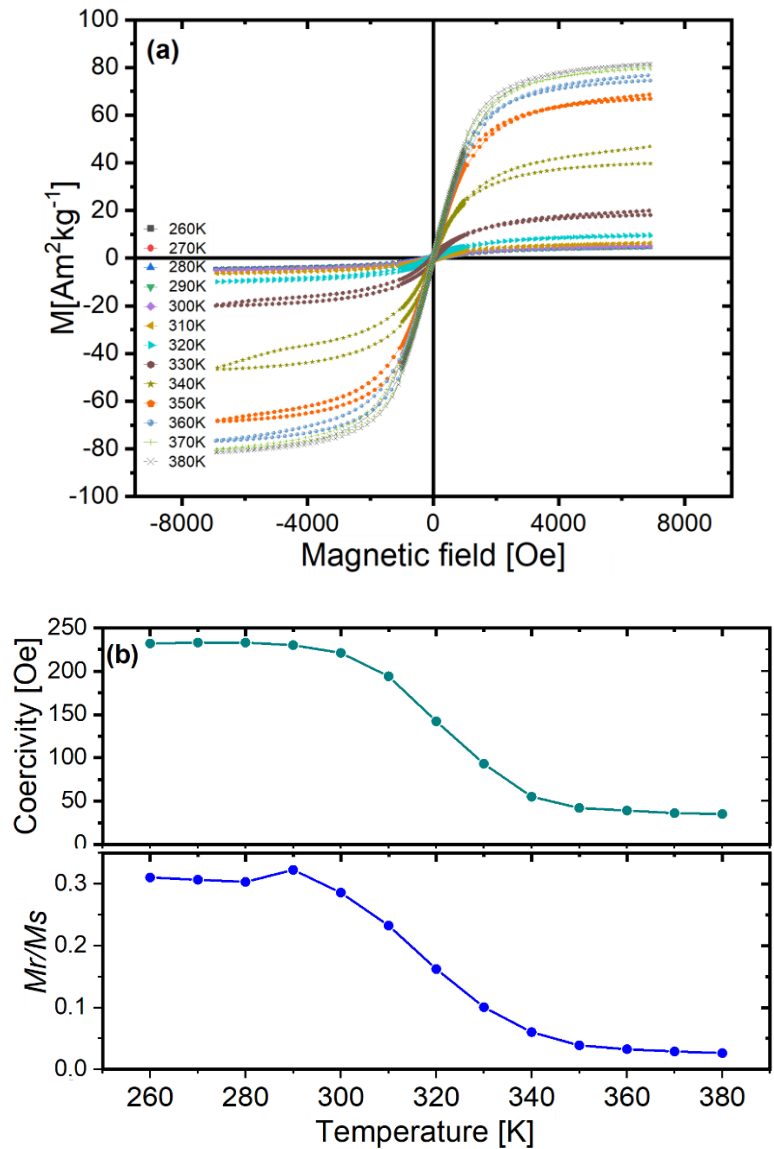

*Figure 3. Magnetic hysteresis loops for the $Fe_{48}Rh_{52}$ sample measured at different temperatures in the heating mode (a); temperature dependence of the coercive force and the squareness coefficient Mr/Ms (b).*

It should be paid attention to the temperature dependence of the squareness coefficient of the hysteresis loop. The general type of dependence is similar to the temperature dependence of coercivity. However, with an increase in temperature from 280 to 290 K, a slight increase in the squareness coefficient is observed. A similar feature was observed in other works [27]. Presumably, this feature is due to the fact that at a given temperature, most of the FM clusters exist in a state close to single-domain. A decrease in the squareness coefficient may be associated with an increase in the number of domains in the FM regions, and therefore with an increase in the FM phase region itself. This result is an additional reason to believe that the micromagnetic structure of the image and its inherent dipole-dipole interactions play a significant role in the evolution of the phase transition, as described in [30].

Application of uniaxial tension and a magnetic field, the general change in entropy, can be represented as

$$\Delta S(T, 0 \to H, 0 \to \sigma) = \Delta S(T, 0, 0 \to \sigma) + \Delta S(T, 0 \to H, \sigma) = \Delta S(T, 0, 0 \to \sigma) + \Delta S(T, 0 \to H, 0) + \int_0^\sigma \int_0^H \frac{\partial}{\partial T}\left(\frac{\partial M}{\partial \sigma}\right)_{T,H} d\sigma dH , \qquad (1)$$

As seen from 1, the third term is responsible for the contribution from cross-effects [12].